\documentclass[twocolumn,secnumarabic,amssymb,superscriptaddress,nofootinbib,nobibnotes, aps, prd]{revtex4-1}

\usepackage[colorlinks, citecolor=blue]{hyperref}
\usepackage[fleqn]{amsmath}
\usepackage{graphics}
\usepackage{xcolor}
\usepackage{bm}
\usepackage{extarrows}
\usepackage{mathrsfs}
\usepackage{empheq}

\def\rar{\rightarrow}

\def\ra{\rangle}
\def\la{\langle}

\def\no{\nonumber}
\def\bea{\begin{eqnarray}}
\def\eea{\end{eqnarray}}
\def\be{\begin{equation}}
\def\ee{\end{equation}}

\def\p{\partial}

\begin{document}

\title{Bell inequalities violation within non-Bunch-Davies states}%

\author{Jun Feng}%
\email{j.feng@xjtu.edu.cn}
\affiliation{School of Science, Xi'an Jiaotong University, Xi'an, Shaanxi 710049, China}
\affiliation{School of Mathematics and Physics, The University of Queensland, Brisbane, QLD 4072, Australia}

\author{Xiaoyang Huang}
\affiliation{School of Science, Xi'an Jiaotong University, Xi'an, Shaanxi 710049, China}

\author{Yao-Zhong Zhang}%
\affiliation{School of Mathematics and Physics, The University of Queensland, Brisbane, QLD 4072, Australia}
\affiliation{Institute of Modern Physics, Northwest University, Xi'an, Shaanxi 710069, China}

\author{Heng Fan}%
\affiliation{Beijing National Laboratory for Condensed Matter Physics, Institute of Physics, Chinese Academy of Sciences, Beijing 100190, China}
\date{\today}%

\begin{abstract}

We study the quantum nature of non-Bunch-Davies states in de Sitter space by evaluating CHSH inequality on a localized two-atom system. We show that quantum nonlocality can be generated through the Markovian evolution of two-atom, witnessed by a violation of CHSH inequality on its final equilibrium state. We find that the upper bound of inequality violation is determined by different choices of de Sitter-invariant vacua sectors. In particular, with growing Gibbons-Hawking temperature, the CHSH bound degrades monotonously for Bunch-Davies vacuum sector. Due to the intrinsic correlation of non-Bunch-Davies vacua, we find that the related violation of inequality can however drastically increase after certain turning point, and may persist for arbitrarily large environment decoherence. This implies that the CHSH inequality is useful to classify the initial quantum state of the Universe. Finally, we clarify that the witnessed intrinsic correlation of non-Bunch-Davies vacua can be utilized for quantum information applications, e.g., surpassing the Heisenberg uncertainty bound of quantum measurement in de Sitter space.

\end{abstract}

\pacs{04.62.+v, 98.80.Cq, 03.65.Ud, 03.65.Yz}

\maketitle

\section{Introduction}
\label{1}

Local measurements performed by distant observers on spacelike-separated quantum systems may lead to nonlocal correlation, which Einstein famously called ``spooky action at a distance.'' This quantum nonlocality has been experimentally revealed by the violation of Bell-type inequalities \cite{Bell1,Bell2}, which places an upper bound on the correlation compatible with classical local hidden-variable theories. While quantum nonlocality completely characterizes the nature of entanglement for all pure states \cite{Bell5}, they are however inequivalent for general mixed states \cite{Bell7}. Indeed, quantum nonlocality provides a profound quantum witness on non-classical correlations \cite{Bell4}, that can be utilized in practical quantum information tasks.

Quantum nonlocality plays an important role in cosmology. In inflationary paradigm, quantum fluctuations would be stretched over to leave a signature as cosmic microwave background (CMB) anisotropies. As the initial state of inflation is heavily squeezed \cite{Bell11}, Bell inequality violating experiment was suggested \cite{Bell12} to be performed on CMB to give compelling evidence for the quantum origin of primordial density fluctuations. Conventionally, quantum fluctuations are assumed to start in a Bunch-Davies (BD) vacuum at infinite past. However, since the field modes below the Planck-scale $\Lambda$ are inaccessible, this simple picture has been seriously questioned in recent years \cite{TP1}. Instead, with a short-distance cutoff at Planck-scale, inflation models with non-BD initial conditions have been extensively studied, in which non-BD vacua are manifested in various approaches, such as, general de Sitter-invariant vacua \cite{TP8}, initial entanglement in scalar fields \cite{TP10} or correlated bubble universes \cite{TP11}. The related trans-Planckian frequencies are expected to red-shift down and leave observable imprint \cite{TP12}. To reveal its quantum origin, Bell-type inequalities violating experiment on the trans-Planckian modification of CMB has been explored recently \cite{Bell20+}. Nevertheless, it is interesting to note that most of above schemes are constructed on global modes of quantum fluctuation, while the cutting-edge cosmic Bell test \cite{Bell18} can only access \emph{localized} quantum states.

In this paper, we study the quantum nature of non-BD states of de Sitter space by evaluating a specific Bell-type inequality (i.e., CHSH inequality \cite{Bell3}). In particular, we are interested in those so-called $\alpha$-vacua, which is an infinite family of de Sitter-invariant vacuum states labeled by a complex number $\alpha$. They were proposed decades ago \cite{TP8} and have recently been rediscovered in \cite{DS1}. With debates on their sick behavior in the UV \cite{DS2}, $\alpha-$vacua nevertheless serve as plausible candidate for non-BD initial condition of inflation at some finite past \cite{TP3}, which leads to anticipated trans-Planckian modification on CMB. For this to be achieved, an appealing connection between $\alpha$ and high-energy cutoff scale $\Lambda$ should be imposed \cite{TP2}. On the other hand, since every $\alpha$-vacuum can be realized as a state of pair condensation, or, alternatively, as a squeezed state over BD vacuum, the quantum uncertainty of $\alpha-$state is heavily constrained. Therefore, we expect that the associated intrinsic quantum correlation can be witnessed by quantum nonlocality \cite{Bell9,ACHSH}, which means that the amount of violation of CHSH inequality should be $\alpha-$dependent.  

Working in detector-field picture, we study a localized system composed of two Unruh-DeWitt detectors in de Sitter space, each modeled by a freely falling two-level atom. Assuming a weakly interaction with a bath of fluctuating scalar field, the two-atom behaves like an open quantum system, suffering by an environment decoherence attributed to thermality of de Sitter-invariant vacua \cite{OPEN1}, e.g., Gibbons-Hawking effect. We evaluate the CHSH inequality on the final equilibrium state of system, and find that the upper bound of inequality violation is determined by the initial state preparation of system, as well as on different choices of superselection sectors of de Sitter vacua. In particular, begin with initial states having no quantum nonlocality, we show that the Markovian evolution of two-atom can lead to final equilibrium states that violate CHSH inequality. With growing Gibbons-Hawking temperature $T_H$, we find that the CHSH bound degrades monotonously for BD vacuum sector, while for non-BD vacua, the violation of inequality can however drastically increase after certain turning point, and may persist for arbitrarily large environment decoherence. This implies that the CHSH inequality is useful to classify the initial quantum state of the Universe \cite{Bell20+}. Moreover, we emphasize that such intrinsic correlation of non-BD vacua witnessed by quantum nonlocality, can be utilized as a physical resource in practical quantum information tasks, such as surpassing the Heisenberg uncertainty bound of quantum measurement in de Sitter space.

The paper is organized as follows. In Section \ref{2}, we solve the master equation of two-atom system, and give explicitly its final equilibrium state. In Section \ref{3}, we evaluate the CHSH inequality for different superselection sectors of de Sitter vacua, and show that the quantum nonlocality arising from intrinsic correlation of non-BD vacua can be utilized as a physical resource in quantum measurement. In Section \ref{4}, the summary and discussion are given. Throughout this paper, we use units with $G=c=\hbar=k_B=1$.

\section{Dynamics of two Unruh-DeWitt detectors}
\label{2}

To proceed CHSH inequality test on Unruh-DeWitt detectors (labeled as $A$ and $B$), we should first explore the full dynamics of bipartite detectors  state in de Sitter background, which effectively is governed by a Lindblad master equation.

\subsection{Master equation of detectors}

Without loss of generality, the total Hamiltonian of the combined system (detectors + environment) is
\be
H_{tot}=H_S+H_\Phi+H_I\label{op1}
\ee
Here $H_S$ is the Hamiltonian of two mutually independent atoms in the common comoving frame. Since each atom internal dynamics is driven by a $2\times2$ matrix, we choose $H_S$ in a concise form as 
\be
H_{S}=\frac{\omega}{2}\bigg(\sigma_3^{(A)}\otimes\mathbf{1}^{(B)}+\mathbf{1}^{(A)}\otimes\sigma_3^{(B)}\bigg)\equiv\frac{\omega}{2}\Sigma_3\label{op1}
\ee
where the symmetrized bipartite operators $\Sigma_i\equiv\sigma_i^{(A)}\otimes\mathbf{1}^{(B)}+\mathbf{1}^{(A)}\otimes\sigma_i^{(B)}$ are defined by Pauli matrices $\sigma_i^{(\alpha)}$ ($i=1,2,3$), with superscript $\alpha=\{A,B\}$ labeling distinct atoms, and $\omega$ represents the energy level spacing of atom. $H_\Phi$ is the Hamiltonian of free massless scalar fields $\Phi(x)$ satisfying Klein-Gordon equation $\Box\Phi(x)=0$ in de Sitter space, with covariant d'Alembertian operator $\Box\equiv g^{\mu\nu}\nabla_\mu\nabla_\nu$ determined by chosen coordinate system. $H_I$ describes the interaction between atoms and scalar field, assumed to be in a form of electric dipole interaction
\be
H_I=\eta~ \Big[(\sigma_2^{(A)}\otimes\mathbf{1}^{(B)})\Phi(t,\mathbf{x}^{(A)})+(\mathbf{1}^{(A)}\otimes\sigma_2^{(B)})\Phi(t,\mathbf{ x}^{(B)})\Big]
\ee
where $\eta$ is a small dimensionless coupling constant. 

We study  the dynamic evolution of detectors' density matrix $\rho_{AB}(t)=\mbox{Tr}_{\Phi}\rho_{tot}(t)$, where $t$ is the proper time of detectors' worldline. We assume that the initial density matrix of total system is separable, i.e., $\rho_{tot} = \rho_{AB}(0) \otimes |0\ra\la0|$, where $|0\ra$ is the vacuum state of field $\Phi(x)$. The total density matrix evolves according to von Neumann equation $i\p_t\rho_{tot}(t) =[H_{tot}(t),\rho_{tot}(t)]$. In a weak coupling limit, the Markovian dynamics of two-atom's density matrix $\rho_{AB}(t)$ can be induced from $\rho_{tot}(t)$, by tracing over all field degrees of freedom, and satisfies a master equation in Kossakowski-Lindblad form \cite{OPEN6}
\be
\frac{\partial\rho_{AB}(t)}{\partial t}=-i[H_{\tiny\mbox{eff}},\rho_{AB}(t) ]+\mathcal{L}[\rho_{AB}(t)]
\label{master}
\ee
where 
\be
\mathcal{L}[\rho_{AB}]=\sum_{\substack{i,j=1,2,3\\   \alpha,\;\beta=A,B}}\frac{C^{(\alpha\beta)}_{ij}}{2}\bigg[2\sigma_j^{(\beta)}\rho_{AB}\sigma_i^{(\alpha)}-\{\sigma_i^{(\alpha)}\sigma_j^{(\beta)},\rho_{AB}\}\bigg]\label{op2}
\ee
is a nonunitary evolution term produced by the coupling with external fields. The Kossakowski matrices $C^{(\alpha\beta)}_{ij}$ can be determined by Fourier transforms of following Wightman functions of scalar field 
\be
G^{(\alpha\beta)}(t-t')=\la0|\Phi(t,\bm{x}^{(\alpha)})\Phi(t',\bm{x}^{(\beta)})|0\ra
\ee
which are
\be
\mathcal{G}^{(\alpha\beta)}(\omega)=\int_{-\infty}^{\infty}d\Delta t~e^{i\omega\Delta t}G^{(\alpha\beta)}(\Delta t)
\ee
where the superscript $\alpha,\beta=\{A,B\}$ labeling distinct atoms. For two-atom system, one can easily find that $G^{(AA)}=G^{(BB)}$ and $G^{(AB)}=G^{(BA)}$, which lead to $\mathcal{G}^{(AA)}=\mathcal{G}^{(BB)}\equiv\mathcal{G}_0$ and $\mathcal{G}^{(AB)}=\mathcal{G}^{(BA)}$.

The master equation (\ref{master}) enables us to describe the asymptotic equilibrium states of detectors at large times, which are governed by the competition between environment dissipation on curved background and quantum correlation generated through the Markovian dynamics of detectors \cite{OPEN5}. For two-atom system, the initial interatomic separation $L\equiv|\bm{x}^{(A)}-\bm{x}^{(B)}|$ is a control parameter of correlation generation, as Kossakowski matrices now become distance-dependent since in general $\mathcal{G}^{(AB)}=\mathcal{G}^{(BA)}\equiv\mathcal{G}(\omega,L)=\mathcal{G}_0(\omega)f(\omega,L)$ for two separated atoms \cite{OPEN1}, where $f(\omega,L)$ is an even function of frequency $\omega$. One would not be surprised \cite{OPEN+1} that the correlation generation between atoms would be more effective for smaller $L$, and becomes impossible for an infinitely large separation. In fact, it was shown \cite{OPEN+3} that there always exist a proper $L$, below which the generated correlation can persist asymptotically in final equilibrium states under environment dissipation. Therefore, we can concisely fix a small interatomic separation, and only concern about the influence of environment decoherence on the equilibrium states of detectors. In such a situation, all the Kossakowski matrices become equal $C^{AA}_{ij}=C^{BB}_{ij}=C^{AB}_{ij}=C^{BA}_{ij}\equiv C_{ij}$ \cite{OPEN+4}, where
\be
C_{ij}=\Gamma_+\delta_{ij}-i\Gamma_-\epsilon_{ijk} \delta_{3k}+\Gamma\delta_{3i}\delta_{3j}\label{op3}
\ee
with
\be
\Gamma_\pm=\frac{1}{2}\big[\mathcal{G}_0(\omega)\pm\mathcal{G}_0(-\omega)\big],\quad \Gamma=\mathcal{G}_0(0)-\Gamma_+\label{op4}
\ee

By resolving the master equation (\ref{master}), the final reduced density matrix of two-atom at asymptotic equilibrium can be expressed in a Bloch form as \cite{OPEN5}
\be
\rho_{AB}(t)=\frac{1}{4}\bigg(\mathbf{1}^{(A)}\otimes\mathbf{1}^{(B)}+\sum_{i=1}^3\rho_{i}\Sigma_i+\sum_{i,j=1}^3\rho_{ij}\sigma_i^{(A)}\otimes\sigma_j^{(B)}\bigg)
\label{op5}
\ee
where
\bea
\rho_{i}&=&-\frac{R}{3+R^2}(\tau+3)\delta_{3i},\no\\
\rho_{ij}&=&\frac{1}{3+R^2}[R^2(\tau+3)\delta_{3i}\delta_{3j}+(\tau-R^2)\delta_{ij}]\label{op6}
\eea
where the ratio $R=\Gamma_-/\Gamma_+$ is determined by the dynamics of the system. The final equilibrium state is also depend on the choice of initial state by $\tau=\sum_i\rho_{ii}(0)$ which is a constant of motion and satisfies $-3\leqslant\tau\leqslant1$ to keep $\rho_{AB}(0)$ positive.

\subsection{Dynamics within general non-BD vacua}
We consider freely falling Unruh-DeWitt detectors in de Sitter space, which weakly interact with a massless scalar field conformally coupling to background. Working with the global coordinate system $(\tilde{t},\chi,\theta,\varphi)$ in which detectors are comoving with the expansion, the line element of de Sitter space becomes
\be
ds^2=d\tilde{t}^2-H^{-2}\cosh^2(H\tilde{t})[d\chi^2+\sin^2\chi(d\theta^2+\sin^2\theta d\phi^2)]\label{cood}
\ee
where Hubble parameter $H$ sets a positive cosmological constant $\Lambda=3H^2$ and curvature radius as $\ell=H^{-1}$. We assume the initial atom separation is sufficiently small, i.e., $L\ll\ell$, so that the generated correlation at asymptotic equilibrium is independent on $L$\footnote{It was claimed in \cite{OPEN1} that quantum correlation generated in de Sitter space cannot be perceived beyond the horizon scale $\ell$, no matter how small the initial separation $L$ is. Nevertheless, once relaxing to master equation with dynamical coarse graining \cite{OPEN+2}, the generated correlation can even survive at the super-horizon scale $>\ell$ if $L$ is small enough. In this paper, we will not further touch on this subtlety.}. Thus, the final equilibrium states of detectors have the form as (\ref{op5}) and (\ref{op6}).

\begin{figure}[hbtp]
\begin{center}
\includegraphics[width=.4\textwidth]{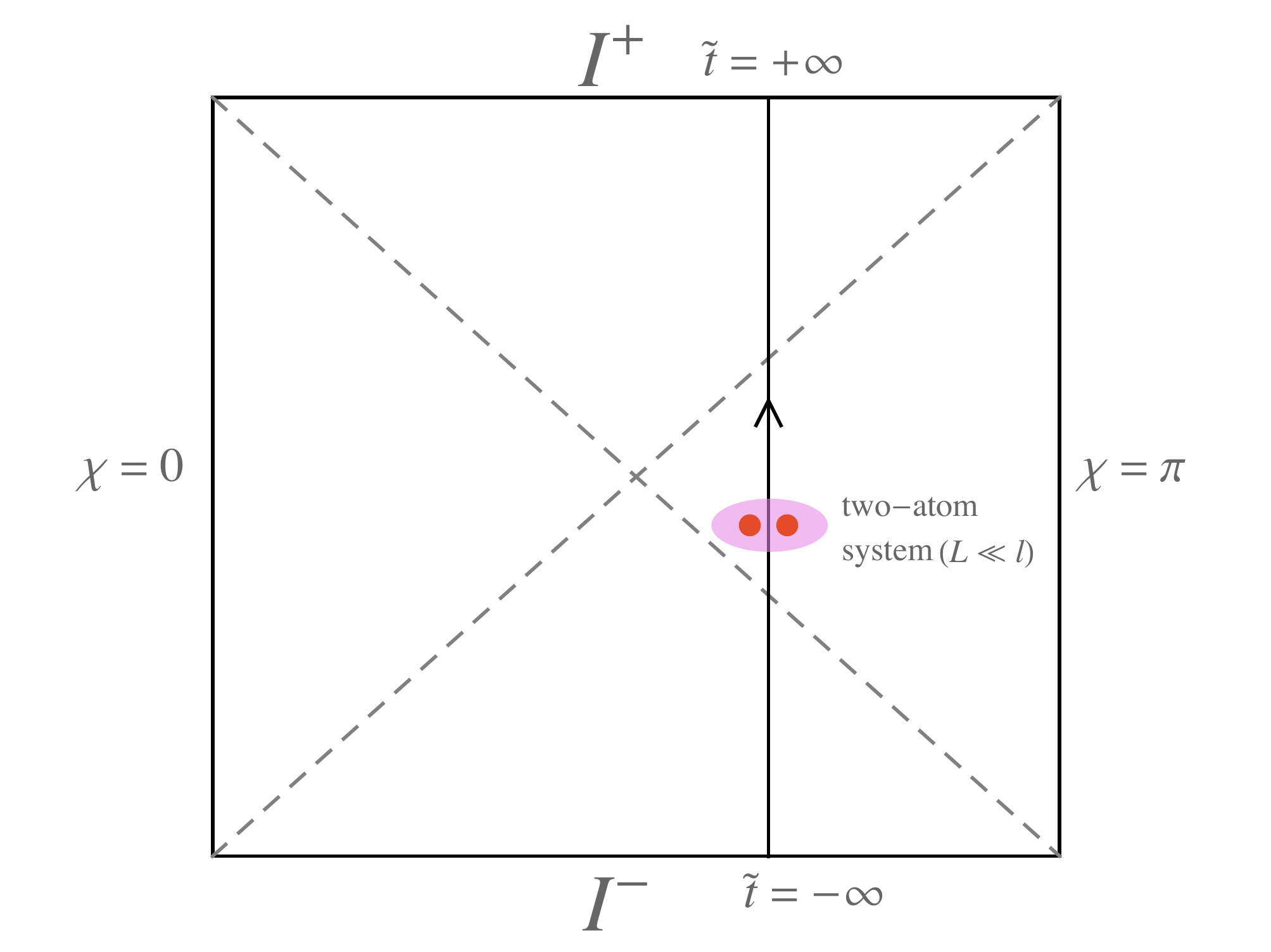}
\caption{Penrose diagram for global de Sitter space in coordinates (\ref{cood}). Two freely falling atoms with small interatomic separation ($L\ll\ell$) move along the geodesic in de Sitter space.}
\label{fig0}
\end{center}
\end{figure}

The scalar wave equation can then be resolved in coordinates (\ref{cood}) and defines a de Sitter-invariant Bunch-Davies vacuum $|BD\ra$. In the massless, conformal coupling limit, for a freely falling detector, the associated Wightman function $G^+_{BD}(x(t),y(t'))\equiv\la BD|\Phi(x(t))\Phi(y(t'))|BD\ra$ becomes \cite{DS+1}
\be
G^+_{BD}(x(t),y(t'))=-\frac{H^2}{16\pi^2\sinh^2[(t-t')H/2-i\epsilon]}\label{Wit1}
\ee
which fulfills KMS condition, i.e., $G^+_{BD}(t)=G^+_{BD}(t+i\beta)$, implying that for a detector in BD vacuum sector, it ends up in thermal equilibrium at universal Gibbons-Hawking temperature $T_H\equiv1/\beta=1/2\pi\ell$ \cite{DS3}.

By Mottola-Allen (MA) transformation, one can further define a one-parameter family of de Sitter-invariant vacua called $\alpha$-vacua $|\alpha\ra$, each of which can be interpreted as a squeezed state over $|BD\ra$, i.e., $|\alpha\ra=\hat{S}(\alpha)|BD\ra$, where $\mbox{Re} \alpha<0$ and $\hat{S}(\alpha)$ denotes a squeezing operator in quantum optics \cite{DS2}. In particular, we adopt CPT invariant $\alpha$-vacua, which means $\alpha<0$ is real. Therefore, the Wightman function for the scalar field in $\alpha-$vacua can be expressed in terms of $G^+_{BD}$ as
\bea
G^+_\alpha(x,y)&=&N^{-1}\Big[G^+_{BD}(x,y)+e^{2\alpha}G^+_{BD}(y,x)\no\\
&&+e^{\alpha}\Big(G^+_{BD}(x,y_A)+ G^+_{BD}(x_A,y)\Big)\Big]\label{op7}
\eea
where subscript $\alpha$ denotes a particular choice of $\alpha$ value, $N\equiv1-e^{2\alpha}$ and $x_A$ is the antipodal point of $x$. As $\alpha\rar-\infty$, we find that $G^+_\alpha(x,y)$ reduces to the Wightman function in BD vacuum $G^+_{BD}(x,y)$, which uniquely extrapolates to the same short-distance behavior of two-point correlation function in the Minkowski vacuum, as the curvature of de Sitter vanishing. 

Employing the relation \cite{OPEN7} 
\be
G^+(x,y_A)=G^+(x_A,y)=G^+(t-i\pi)
\ee
and substituting (\ref{Wit1}) into (\ref{op7}), we can calculate the Fourier transformation of Wightman function in non-BD sectors as
\be
\mathcal{G}_\alpha=\frac{\omega(1+e^{\alpha-\pi\omega})^2}{2\pi(1-e^{-\beta\omega})(1-e^{2\alpha})}
\ee
and the related Kossakowski coefficients are 
\bea
\Gamma_{+,~\alpha}&=&\frac{\omega\Big[(1+e^{\alpha-\pi\omega})^2+e^{-\beta\omega}(1+e^{\alpha+\pi\omega})^2\Big]}{4\pi(1-e^{2\alpha})(1-e^{-\beta\omega})}\no\\
\Gamma_{-,~\alpha}&=&\frac{\omega\Big[(1+e^{\alpha-\pi\omega})^2-e^{-\beta\omega}(1+e^{\alpha+\pi\omega})^2\Big]}{4\pi(1-e^{2\alpha})(1-e^{-\beta\omega})}\no\\
R_\alpha&=&\frac{(1+e^{\alpha-\pi\omega})^2-e^{-\beta\omega}(1+e^{\alpha+\pi\omega})^2}{(1+e^{\alpha-\pi\omega})^2+e^{-\beta\omega}(1+e^{\alpha+\pi\omega})^2}\label{op8}
\eea
For fixed energy level spacing of detectors, it is easy to find that $R_\alpha\in[-1,+1]$ as $\beta$ varying. Inserting (\ref{op8}) into (\ref{op6}), we obtain the final equilibrium state of two detectors respecting to $\alpha-$vacua, which is
\be
\rho_{AB}
=\left(\begin{array}{cccc}
A_- & 0 & 0 & 0 \\
0 & B & C & 0 \\
0 & C & B & 0 \\
0 & 0 & 0 & A_+\end{array}\right)
\label{op9}
\ee  
where
\bea
A_\pm&=&\frac{(3+\tau)(R_\alpha\pm1)^2}{4(3+R_\alpha^2)},~~B=\frac{3-\tau-(\tau+1)R_\alpha^2}{4(3+R_\alpha^2)},\no\\
C&=&\frac{\tau-R_\alpha^2}{2(3+R_\alpha^2)}\no
\eea

Before moving to the quantum nonlocality of our model, several remarks on $\alpha-$vacua should be addressed. Firstly, the squeezing nature of $\alpha-$vacua can heavily constrain the measurement uncertainty. This implies that certain intrinsic quantum correlation could be concealed in these non-BD states \cite{Bell9}. Secondly, $\alpha-$vacua are not thermal in character, following the fact that (\ref{op7}) fails KMS condition unless it reduces to BD vacuum. Such deviation from thermality has attracted many attempts \cite{TP3} to take $\alpha-$vacua as alternative initial state of inflation. To match the anticipated correction in the primordial power spectrum of order $\sim \mathcal{O}(H/\Lambda)^2$, the parameter $\alpha$ can directly be connected to $\Lambda$, some fundamental scales of new physics (e.g., the Planck scale or the stringy scale) \cite{TP2}. 

\section{Quantum nonlocality in general de Sitter vacua}
\label{3}

To explore the quantum nature of intrinsic correlation within non-BD vacua, we perform a Bell-type inequalities violating experiment on the bipartite state of Unruh-DeWitt detectors. We expect that the amount of violation should depend on initial state preparation of detectors and different choices of superselection sectors of $\alpha-$vacua. 

\subsection{Violation of Bell inequality in de Sitter-invariant vacua}

We would check that if the reduced density matrix (\ref{op9}) can violate CHSH inequality, which is a suitable inequality for two qubits to test local-realistic theories. Any such theory must satisfy the bound \cite{Bell3}
\be
|\la\,\mathcal{B}_{CHSH}\,\ra_{\rho}|\,\leq\,2
\label{op10}
\ee
where the CHSH expectation value for a given state $\rho$ is defined as $\mathcal{B}_{CHSH}=\mathbf{a}\cdot\bm{\sigma}\otimes(\mathbf{b}+\mathbf{b^{\prime}})\cdot\bm{\sigma}+\mathbf{a^{\prime}}\cdot\bm{\sigma}\otimes(\mathbf{b}-\mathbf{b^{\prime}})\cdot\bm{\sigma}$, and $\bm{\sigma}$ is the vector of Pauli matrices. Therefore, the nonlocal correlation of a quantum state $\rho$ can be witnessed by the violation of (\ref{op10}) up to a value $2\sqrt2$, for some choices of unit vectors $\mathbf{a}$, $\mathbf{b}$, $\mathbf{a^{\prime}}$ and $\mathbf{b^{\prime}}$ in $\mathbb{R}^{3}$.  

For general bipartite states, a criterion for the maximally possible violation of CHSH inequality has been proved in \cite{Bell21}, which claims that the CHSH bound, i.e., the maximal possible value $\la\,\mathcal{B}_{CHSH}\,\ra_{\rho}$ can be determined by
\be
\la\,\mathcal{B}_{max}\,\ra_{\rho}\,=\,2\,\sqrt{\nu_{1}+\nu_{2}}
\label{op11}
\ee
Here $\nu_{1},\nu_{2}$ are two largest eigenvalues of matrix $T^{t}_{\rho}T_{\rho}$, where the components of correlation matrix $T_{\rho}=(t_{ij})$ are defined as $t_{ij}=\mbox{Tr}[\rho\,\sigma_{i}\otimes\sigma_{j}]$. 

%
%

In our case, we can straightforwardly obtain a diagonal $T^{t}_{\rho_{AB}}T_{\rho_{AB}}$ matrix from (\ref{op9}) with three eigenvalues
\be
\lambda_1=\lambda_2=\Bigg(\frac{\tau-R_\alpha^2}{3+R_\alpha^2}\Bigg)^2~~,~~\lambda_3=\Bigg[\frac{\tau+(\tau+2)R_\alpha^2}{3+R_\alpha^2}\Bigg]^2\label{op12}
\ee
which indicates that $\la\,\mathcal{B}_{max}\,\ra_{\rho_{AB}}$ should depend on the initial state of detectors labeling by $\tau$, and the superselection sectors of vacua labeling by $\alpha$. According to different preparations of initial state $\tau$, we come to three classes of maximal amount of CHSH inequality violation:

\textbf{(i)} For the initial state prepared with $\tau_{0}=-(1+2/R_\alpha^2)^{-1}$, we have $\lambda_1=\lambda_2=\lambda_3$. The CHSH bound (\ref{op11}) is
\be
\la\,\mathcal{B}_{max}\,\ra_{\scriptsize\mbox{(i)}}=2\sqrt2|\tau_0|=\frac{2\sqrt2R_\alpha^2}{2+R_\alpha^2}\leqslant\frac{2\sqrt2}{3}  \label{op13}
\ee
since $R_\alpha\in[-1,+1]$. No violation of CHSH inequality for final equilibrium state can be found as the CHSH bound (\ref{op13}) is $\sim0.9428$, which means that state (\ref{op9}) is local for any superselection sectors of non-BD vacua.  

\textbf{(ii)} For the initial state prepared with $\tau\in(\tau_0,1]$, we have $\lambda_1=\lambda_2<\lambda_3$. The corresponding CHSH bound is
\bea
&&\la\,\mathcal{B}_{max}\,\ra_{\scriptsize\mbox{(ii)}}=2\,\sqrt{\lambda_{1}+\lambda_{3}}\no\\
&=&\frac{2}{3+R_\alpha^2}\sqrt{R_\alpha^4(\tau^2+4\tau+5)+2R_\alpha^2(\tau^2+\tau)+2\tau^2}\label{op14}\no\\
\eea
Since $R_\alpha\in[-1,+1]$ for varying $\beta$, we can evaluate (\ref{op14}) and find it cannot across the classical bound of 2, which indicates that there is no quantum nonlocality for final state (\ref{op9}) with such initial state preparation.

\textbf{(iii)} For the initial state preparing with $\tau\in[-3, \tau_0)$, we have $\lambda_1=\lambda_2\geqslant\lambda_3$. Then the CHSH bound becomes
\be
\la\,\mathcal{B}_{max}\,\ra_{\scriptsize\mbox{(iii)}}=2\sqrt{2\lambda_{1}}=2\sqrt2\;\frac{\big|\tau-R_\alpha^2\big|}{3+R_\alpha^2}
\label{op15}
\ee
It is easy to observe that for initial state prepared with $\tau=-3$, the CHSH inequality can be maximally violated for final equilibrium state up to $2\sqrt2$, i.e., so-called Tsirelson bound, which is the upper limit to quantum correlations between distant events. We note that such violation is independent on any choice of $\alpha-$vacua. 

For general $\tau\in(-3, \tau_0)$, things are complicated for (\ref{op15}) by its dependence on the infinite family of non-BD vacua. In the following, we want to show that for some initial states satisfying inequality (\ref{op10}), quantum nonlocality can be generated in final state  (\ref{op9}) after Markovian evolution of the system, and be witnessed by violating CHSH inequality in de Sitter space. 

To illustrate this, we consider two Unruh-DeWitt detectors initially prepared in a Bell-diagonal state in a freely falling basis
\be
\rho_0=\frac{1}{4}\bigg(\mathbf{1}^{(A)}\otimes\mathbf{1}^{(B)}+\sum_{i=1}^{3}c_i\sigma_i^{(A)}\otimes\sigma_i^{(B)}\bigg)
\label{op16}
\ee
with coefficients $0\leqslant|c_i|\leqslant1$. These states are the convex combination of four Bell states and reduce to maximally entangled states (Bell-basis) if $|c_1|=|c_2|=|c_3|=1$. The state (\ref{op16}) is a generalization of many important quantum states, e.g., Werner state \cite{Bell7}, and has important applications in quantum information. 

\begin{figure}[hbtp]
\begin{center}
\includegraphics[width=.48\textwidth]{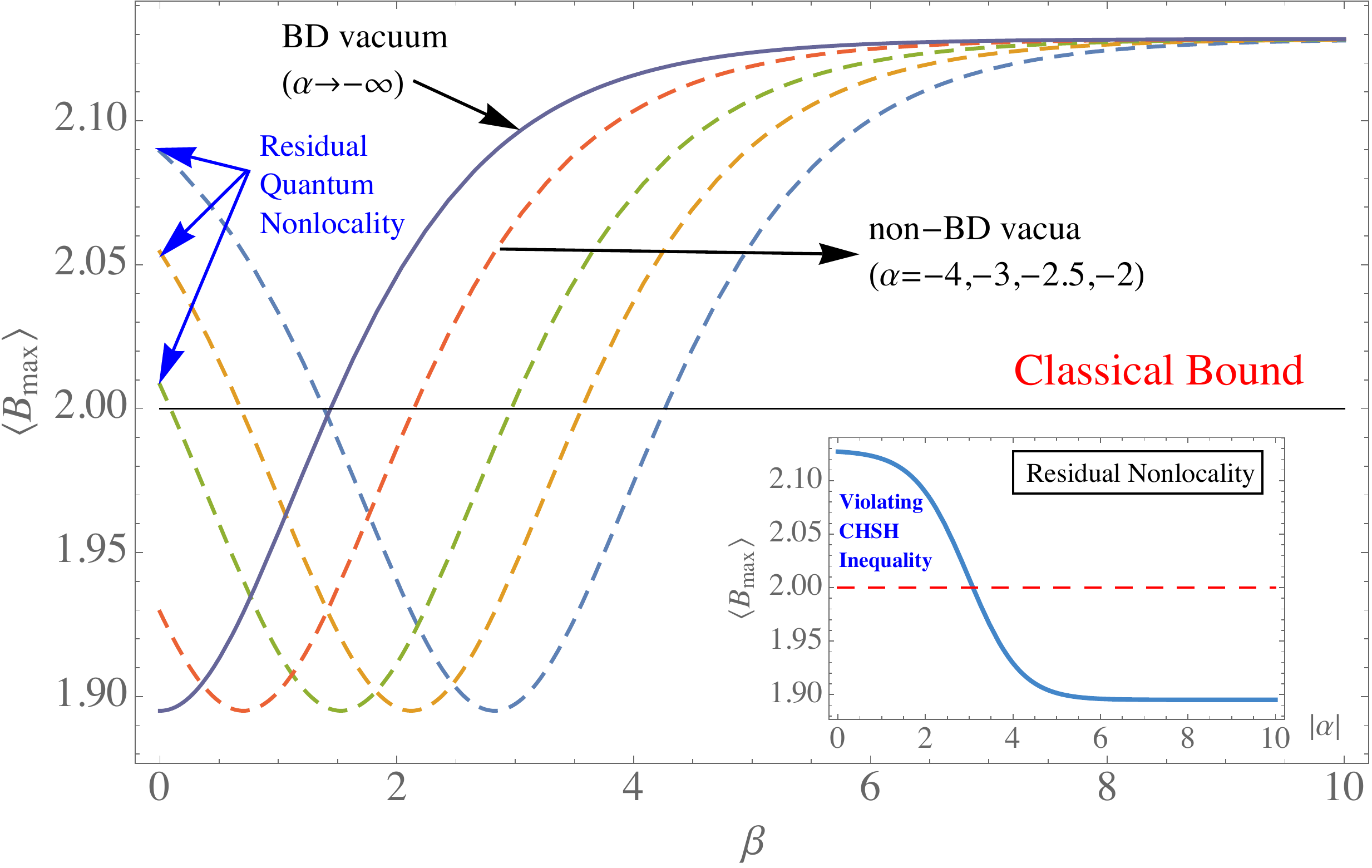}
\caption{Violation of CHSH inequality. The initial state of detectors is chosen as Bell-diagonal state with $c_1=c_2=c_3=-2/3$ (i.e., $\tau=-2$), which satisfies CHSH inequality. With quantum correlation generated through Markovian evolution of the system, the inequality can be violated for the final state of detectors. For general $\alpha-$vacua, the amount of CHSH bound drastically increases after certain turning point $\beta_c$. In particular, for vacua sectors with $\alpha=-2,-2.5,-3$, the residual quantum nonlocality can survive at arbitrarily large Gibbons-Hawking temperature. In the inset, the $\alpha$ dependence of residual quantum nonlocality at $\beta\rar0$ is demonstrated, where a bigger violation of CHSH inequality happens for larger deviation from BD superselection sector.}
\label{fig1}
\end{center}
\end{figure}

By performing a CHSH violating experiment on the initial states (\ref{op16}), we can determine the related CHSH bound. The matrix $T^{t}_{\rho_{0}}T_{\rho_{0}}$ is diagonal with three eigenvalues $c_1^2,c_2^2,c_3^2$. We consider an instructive example with $c_1=c_2=c_3=-2/3$, which gives CHSH bound $\la\,\mathcal{B}_{max}\,\ra_{0}=2\sqrt2|c_1|\approx1.8856$, indicating that the bipartite detectors have no quantum nonlocality initially. As $\tau=-2$ in this case, the CHSH bound (\ref{op15}) gives 
\be
\la\,\mathcal{B}_{max}\,\ra_{\scriptsize\mbox{Bell-diag}}=\frac{1}{\sqrt2}\frac{3\gamma_-^2+3e^{-2\beta\omega}\gamma_+^2+2\gamma_+\gamma_-e^{-\beta\omega}}{\gamma_-^2+e^{-2\beta\omega}\gamma_+^2+\gamma_+\gamma_-e^{-\beta\omega}}
\label{op17}
\ee
where $\gamma_\pm\equiv(1+e^{\alpha\pm\pi\omega})^2$. We depict (\ref{op17}) in Fig.\ref{fig1} for various $\alpha$, corresponding to different superselection sectors of de Sitter vacua.

We find that although the initial Bell-diagonal state is chosen without quantum nonlocality, the final state of detectors can violate the CHSH inequality, which implies that the quantum correlation has been generated through Markovian evolution of system and overcome Gibbons-Hawking decoherence in de Sitter space. With growing Gibbons-Hawking temperature (i.e., degrading $\beta$), for BD sector of $\alpha\rar-\infty$, the amount of CHSH bound monotonously degrades, consistent with the results of \cite{OPEN1}.

For non-BD sectors of vacua, however, we find that there is a drastically increase of CHSH bound when $\beta$ across certain turning point, determined by $\p_\beta \la\,\mathcal{B}_{max}\,\ra_{\scriptsize\mbox{Bell-diag}}=0$ as
\be
\beta_c(\alpha)=\frac{1}{\omega}\ln\frac{\gamma_+}{\gamma_-}\label{op18}
\ee 
With larger deviation from BD superselection sector, we observe a bigger amount of violation of CHSH inequality for non-BD vacuum sectors. This implies that the CHSH inequality is useful to classify the initial quantum state of the Universe in an inflationary context. Moreover, starting with same initial state, for some $\alpha-$vacua, we find that CHSH bound under arbitrarily large Gibbons-Hawking decoherence (as $\beta\rar0$) can surpass the classical bound, which means that nonvanishing \emph{residual} quantum nonlocality survives\footnote{Strictly speaking, we cannot approach $\beta\rar0$ ($\ell\rar0$) arbitrarily close since new physics around Planckian/stringy scales may enter. The natural way to resolve this is replacing (\ref{Wit1}) by a deformed Wightman function to incorporate Planck cutoff \cite{Planck2}. Nevertheless, in present analysis, we will not discuss details further and just emphasize the persistence of generated quantum nonlocality in non-BD sectors under large Gibbons-Hawking decoherence, distinct from previous findings.}, as depicted in the inset of Fig.\ref{fig1}. It is interesting to note that our result contrasts sharply to previous findings in \cite{ACHSH}, which claimed that no quantum nonlocality can exist in the limit of $\beta\rar0$, regardless of any vacuum sectors. Nevertheless, with in mind that our Bell violating experiment is performed on a local quantum system, we need not surprise on its difference to the Bell test in \cite{ACHSH} which utilizes global field modes in non-BD sectors. 

The increment of CHSH bound in Fig.\ref{fig1} can naturally be attributed to intrinsic quantum correlation of $\alpha-$vacua, since they are constructed by squeezing operator $\hat{S}(\alpha)\sim \exp[\alpha(\hat{a}^{\dag2}-\hat{a}^2)]$ on BD vacuum. For small squeezing parameter $|\alpha|$, the quantum uncertainty of the state can be constrained so heavily, such that the intrinsic correlation grows faster than Gibbons-Hawking decoherence.

\subsection{Quantum nonlocality as physical resource}

In the following, we show that the quantum nonlocality arising from $\alpha-$vacua can be utilized as a physical resource within quantum information applications. In particular, we consider an uncertainty game between Unruh-DeWitt detectors and show that the witnessed intrinsic correlation of non-BD vacua can be used to surpass the Heisenberg uncertainty bound of quantum measurement in de Sitter space.

Firstly, we would like to clarify the inequivalent between quantum nonlocality serving as a witness of quantum correlation and quantum entanglement itself for general $\alpha-$vacua. We choose quantum negativity as a measure of distillable entanglement for detectors' final equilibrium state, defined \cite{NEG} as $\mathcal{N}(\rho)=\frac{1}{2}\sum_i(|\tilde{\lambda}_i|-\tilde{\lambda}_i)=-\sum_{\tilde{\lambda}_i<0}\tilde{\lambda}_i$, where $\tilde{\lambda}_i$ are the negative eigenvalues of partial transposed density matrix of $\rho$. The value of negativity ranges from 0, for separable states, to 0.5, for maximally entangled states. Form (\ref{op9}), we can straightforwardly obtain \cite{FENG}
\be
\mathcal{N}=\frac{2\sqrt{R_\alpha^4+R_\alpha^2(9+4\tau+\tau^2)+\tau^2}-(3+\tau)(1+R_\alpha^2)}{4(3+R_\alpha^2)}
\label{op19}
\ee 
which reaches maximum 0.5 at $\tau=-3$ and becomes vanishing at $\tau=(5R_\alpha^2-3)/(3-R_\alpha^2)$. 

\begin{figure}[hbtp]
\begin{center}
\includegraphics[width=.48\textwidth]{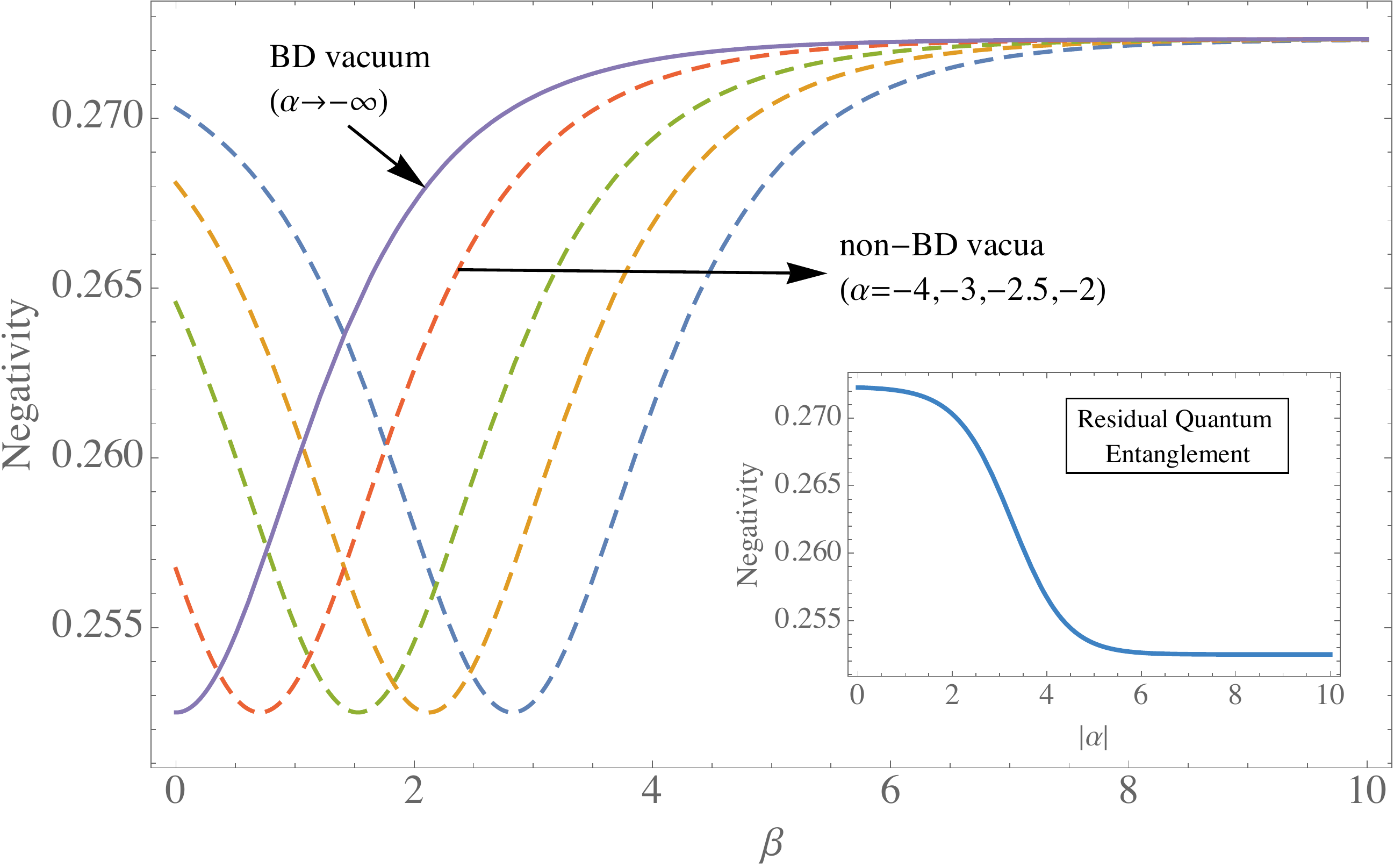}
\caption{Negativity as a measure of entanglement. The initial state of detectors is chosen as Bell-diagonal state with $c_1=c_2=c_3=-2/3$ (i.e., $\tau=-2$). For general $\alpha-$vacua, negativity can drastically increase after turning point $\beta_c$. In the inset, the $\alpha$ dependence of residual quantum entanglement at infinite Gibbons-Hawking temperature is demonstrated, where the larger negativity happens for larger deviation from BD superselection sector.}
\label{fig2}
\end{center}
\end{figure}

For different superselection sectors of vacua, we depict the related negativity in Fig.\ref{fig2}. While entanglement exhibiting a similar dynamics w.r.t. $\beta$ as the CHSH bound in Fig.\ref{fig1}, residual entanglement can survive with arbitrarily large Gibbons-Hawking temperature for every superselection sector of vacua, including BD sector. However, as we showed before, except part of non-BD vacuum sectors (e.g., sectors with $\alpha=-2,-2.5,-3$ in Fig.\ref{fig1}), the final state of detectors cannot violate CHSH inequality, which means that no quantum information tasks using these entanglement can outperform states with appropriate classical correlations. In this meaning, witnesses on quantum correlation such as quantum nonlocality just quantify the part of entanglement that can be utilized as a physical resource in practical quantum information process.

Regarding the squeezing nature of non-BD vacua, their intrinsic correlation can be recast by an entropic form of quantum uncertainty relation \cite{EUR1}. In particular, we consider an uncertainty game played by  Alice ($A$) and Bob ($B$), freely falling two-level atoms in de Sitter space. For incompatible measurements $Q$ and $R$ on quantum system $A$, with $B$ serving as a memory, one arrives at quantum-memory-assisted entropic uncertainty relation (EUR) \cite{EUR2}
\be
S(Q|B)+S(R|B)\geqslant U_b
\label{op20}
\ee
where $U_b\equiv-\log_2c+S(A|B)$ is the entropic uncertainty bound (EUB), and $c=\mbox{max}_{i,j}|\la a_i| b_j\ra|^2$ quantifies the complementarity of observables $Q$ and $R$ with eigenvectors $|a_i\ra$ and $|b_j\ra$. For disentangled $A$ and $B$, the quantum conditional von Neumann entropy $S(A|B)=S(\rho_{AB})-S(\rho_{B})$ satisfies $S(A|B)\geqslant0$, which means that the lower $U_b$ is at most the Heisenberg uncertainty bound $-\log_2c$. However, once $A$ and $B$ are entangled, $S(A|B)$ may possible be negative, leading to a EUB lower than Heisenberg bound. In the extreme case where $A$ and $B$ are maximally entangled, $U_b=0$, enabling us to predict the outcomes precisely. Experimentally, EUB has served \cite{EUR3} as a novel witness of quantum entanglement between $A$ and $B$.  

We would like to compare EUB for (\ref{op9}) with related quantum nonlocality discussed before. For simplicity, we assume that the qubit $A$ of Alice is measured by one of Pauli operators, which gives $c=\frac{1}{2}$ for any two observables $\sigma_j$ and $\sigma_k$ ($j\neq k=1,2,3$). In our case, we apply the EUB of (\ref{op20}) for the final equilibrium state (\ref{op9}), which gives \cite{FENG}
\bea
U_b&=&\frac{3+\tau}{4}\Bigg[\log_2(3+R_\alpha^2)-\sum_{\epsilon=\pm}\bigg(1+\frac{ 4\epsilon R_\alpha}{R^2_\alpha+3}\bigg)\log_2(1+\epsilon R_\alpha)\Bigg]\no\\
&&\no\\
&&+H_{\tiny\mbox{bin}}\bigg(\frac{1-\tau}{4}\bigg)-H_{\tiny\mbox{bin}}\bigg[\frac{1}{2}\bigg(1-\frac{3+\tau}{1+3/R_\alpha}\bigg)\bigg]+1
\label{op21}
\eea
where the binary entropy is defined as $H_{\tiny\mbox{bin}}(p)\equiv-p\log_2p-(1-p)\log_2(1-p)$, and $R_\alpha$ is given by (\ref{op8}). For specific initial state with $\tau$, we expect that the intrinsic correlation of superselection sectors of non-BD vacua can be witnessed by $U_b$ through its violation of Heisenberg uncertainty bound, i.e., $U_b\leqslant1$ in our case.

\begin{figure}[hbtp]
\begin{center}
\includegraphics[width=.48\textwidth]{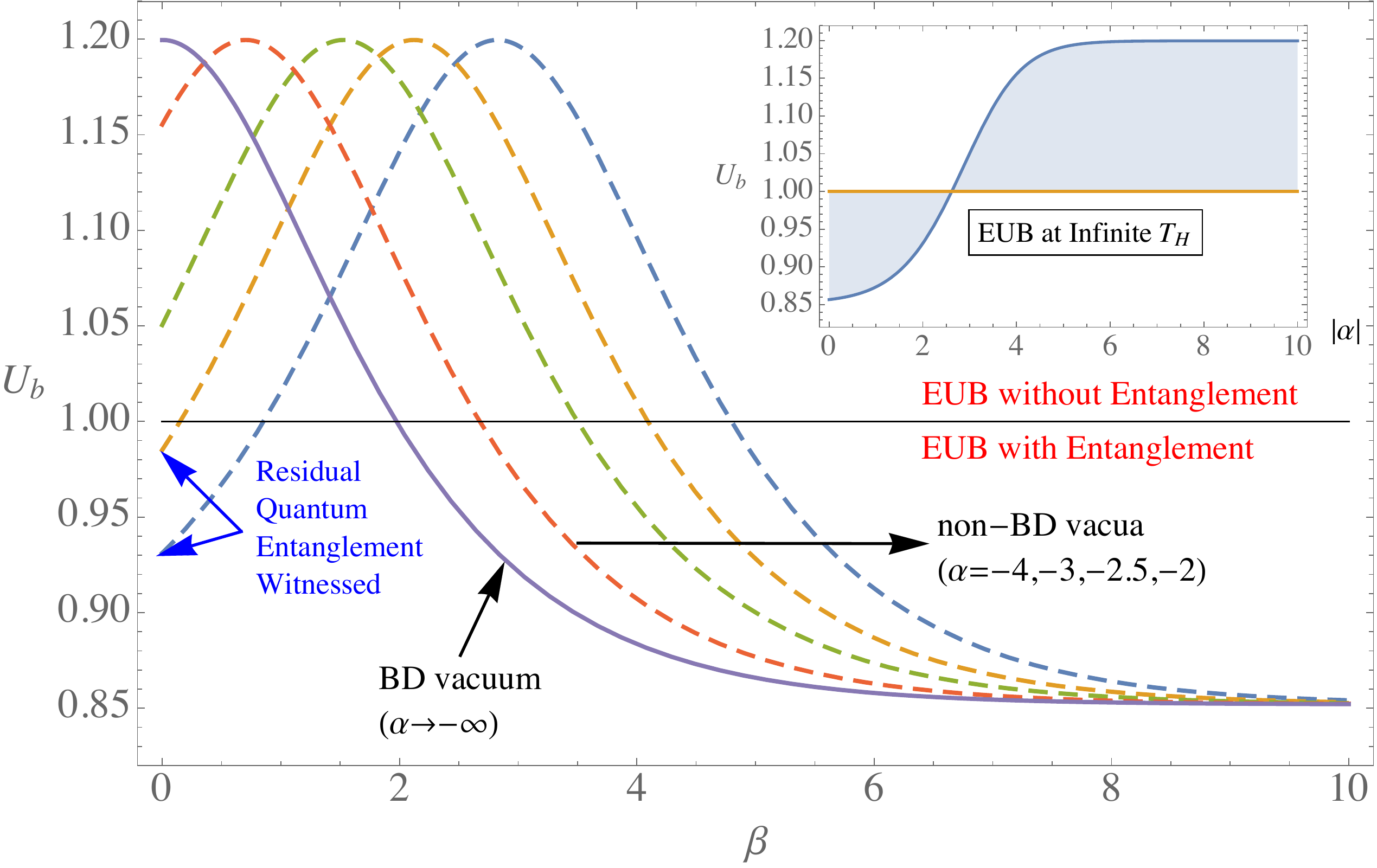}
\caption{EUB as a witness of quantum entanglement. The initial state of detectors is chosen as Bell-diagonal state with $c_1=c_2=c_3=-2/3$ (i.e., $\tau=-2$). For non-BD vacua sectors with $\alpha=-2,-2.5$, the Heisenberg uncertainty bound $U_b=1$ can be violated at arbitrarily large Gibbons-Hawking temperature, witnessing the residual entanglement between detector and quantum memory. In the inset, the $\alpha$ dependence of EUB at $\beta\rar0$ is demonstrated.}
\label{fig3}
\end{center}
\end{figure}

We depict the dynamics of EUB (\ref{op21}) w.r.t. $\beta$ in Fig.\ref{fig3} for superselection sectors of non-BD vacua with $\alpha=-2,-2.5,-3,-4$, as well as BD vacuum sector with $\alpha\rar-\infty$. Firstly, we observe that EUB of quantum measurement performed by detector can violate the Heisenberg bound $U_b=1$ at large $\beta$, but growing with high Gibbons-Hawking temperature, which is not surprising as the thermality of de Sitter vacua should introduce classical noise in quantum measurement \cite{EUR4}. Nevertheless, for non-BD vacuum sectors, $U_b$ degrades once across certain turning point $\beta_c$, consistent with the behavior of CHSH bound in Fig.\ref{fig1}. In particular, we find that for arbitrarily large Gibbons-Hawking temperature, residual quantum correlation for $\alpha=-2,-2.5$ can be utilized to surmount the Heisenberg bound through $U_b<1$, compatible with the interpretation that only the entanglement recognized by quantum witnesses (e.g., violation of CHSH inequality) can be employed in practical quantum information tasks. Moreover, since the residual entanglement at $\beta\rar0$ for $\alpha=-3$ can be recognized by the violation of CHSH inequality (\ref{op17}) but not the EUB (\ref{op21}), we should refer quantum nonlocality as a more profound quantum witness.  

\section{Conclusions}
\label{4}

In this paper, we explore quantum nonlocality as a witness to intrinsic quantum correlation of non-BD vacua in de Sitter space. By performing CHSH inequality violating test on the localized state of two Unruh-DeWitt detectors, we find that the amount of inequality violation quantified by CHSH bound can drastically increase after certain turning point $\beta_c$ for non-BD sectors, as a result of competition between Gibbons-Hawking decoherence and intrinsic correlation of $\alpha-$vacua. Since we found a clear difference between BD vacuum and non-BD vacua in the violation of CHSH inequality, we may be able to classify the initial quantum state in inflationary cosmology. In particular, quantum nonlocality in some non-BD sectors can persist under arbitrarily large Gibbons-Hawking decoherence, contrasting sharply with results of \cite{ACHSH} employing global field modes. Moreover, we show that those quantum correlation witnessed by quantum nonlocality can be utilized as a physical resource in practical quantum information process, such as surpassing Heisenberg uncertainty bound of quantum measurements in de Sitter space.

Our study raises several implications. Firstly, besides attributing to the squeezing nature of $\alpha-$vacua, an alternative explanation on the increment of CHSH bound in Fig.\ref{fig1} may appeal to the connection between parameter $\alpha$ and fundamental scale $\Lambda$ of quantum gravity, e.g., $e^\alpha\sim(H/\Lambda)$ implied by trans-Planckian physics in cosmology \cite{TP3}. The scale cutoff at $\Lambda$ means that we can hardly distinguish the events in de Sitter space once across $\Lambda$ to infinitesimal scales \cite{OPEN8}. Inherited by local quantum state of detectors, such \emph{indistinguishability} of states below scale $\Lambda$ might be manifested in quantum nonlocality. Once we expect an universal relation between $\Lambda$ and vacua selection should exist in de Sitter space, suggested as in trans-Planckian issue of inflation, the quantum nonlocality can be witnessed by an $\alpha-$dependent CHSH bound beyond classical limit. In this meaning, the increment of CHSH bound for general $\alpha$ may imply a possible quantum gravity effect, potentially emerging at the scale determined from (\ref{op18}). Such interpretation of $\beta_c$ as possible manifestation of quantum gravity may further be supported from the study on quantum communication between local events in de Sitter space \cite{ACHSH}, where the quantum capacity of Grassmann communication channel approaches zero at certain $\alpha-$dependent scale. 

Moreover, we can go further to ask that, regardless of specific inflation model, if any invariant Planck scale cutoff in quantum gravity can be witnessed by the emergence of quantum nonlocality. Evidences shown \cite{Planck2} that the delocalization caused by the minimal length should lead to a deformed two-point correlation function, which modifies the spectrum of a detector in curved spacetime. By performing Bell-type inequalities violating experiment on the localized state of detector, the quantum nonlocality arose directly from minimal length is expected to be witnessed. 

Finally, while we have evaluated a Bell's test for detectors in equilibrium state purely determined by the thermality of $\alpha-$vacua, it would be interesting to further explore that how the time-dependence of de Sitter background may be encoded into quantum nonlocality of open quantum system. To achieve this, rather than be placed in an environment of given (non-)BD states, we assume the detectors interact with a scalar field $\Phi(x)$ in its instantaneous ground state $|0_t\ra$ (with $|BD\ra=\lim_{t\rar-\infty}|0_t\ra$) at each instant \cite{master1}. We should note that this leads to a different relaxation process from the one where the detectors get adjusted to their given environment. The later process can be neglected by taking large time limit to give asymptotical equilibrium state as we did in Section \ref{2}. Starting within instantaneous ground state, the corresponding Wightman function of scalar field is then deformed from $G^+_{BD}$ to have additional damping term \cite{master1}, and  becomes time-dependent under Fourier transformation. Following same analysis in this paper, we can eventually deduce a CHSH bound undergoing a time-dependent relaxation inherited from the nonequilibrium dynamics of de Sitter background. We will report the related work elsewhere.

\section*{Acknowledgement}
This work is supported by the National Natural Science Foundation of China (No. 11505133), the Fundamental Research Funds for the Central Universities, Natural Science Basic Research Plan in Shaanxi Province of China (No. 2018JM1049) and the Postdoctoral Science Foundation of China (No. 2016M592769). We thank the referee for constructive comments. J.F. thanks Liu Zhao for stimulating discussions. X.Y.H. acknowledges the support of Innovation and Research Program (No. XJ201710698112). H.F. acknowledges the support of the National Natural Science Foundation of China (No. 91536108, 11774406) and National Key R\&D Program of China (No. 2016YFA0302104, 2016YFA0300600). Y.Z.Z. acknowledges the support of the Australian Research Council Discovery Project (No. 140101492) and the National Natural Science Foundation of China (No. 11775177).

\end{document}